\documentclass[conference]{IEEEtran}
\IEEEoverridecommandlockouts

\usepackage{cite}

\ifCLASSINFOpdf
\else
\fi
\usepackage{amsmath}
\makeatletter
\newcommand{\pushright}[1]{\ifmeasuring@#1\else\omit\hfill$\displaystyle#1$\fi\ignorespaces}
\newcommand{\pushleft}[1]{\ifmeasuring@#1\else\omit$\displaystyle#1$\hfill\fi\ignorespaces}
\makeatother

\usepackage{graphicx}

\usepackage{url}

\usepackage{adjustbox}
\usepackage{multirow}

\begin{document}
%



\title{A Novel Numerical Index for Assessing Results of Frequency Response Analysis (FRA): \\an Experimental Study on Electrical Machines}

\author{
\IEEEauthorblockN{Reza Khalilisenobari\textsuperscript{1} and Javad Sadeh\textsuperscript{*}}
\IEEEauthorblockA{Electrical Engineering Department, Faculty of Engineering, Ferdowsi University of Mashhad, Mashhad, Iran}
\thanks{\textsuperscript{1} Reza Khalilisenobari was with the Department of Electrical Engineering, Ferdowsi University of Mashhad, Mashhad, Iran. Currently, he is with the School of Electrical, Computer and Energy Engineering, Arizona State  University, Tempe, AZ, U.S. Email address: rezakhalili@asu.edu}
\thanks{\textsuperscript{*} Corresponding author at: Electrical Engineering Department, Faculty of Engineering, Ferdowsi University of Mashhad, Mashhad, Iran. Tel.: +98 513 8763302; fax: +98 513 8763302. E-mail address: sadeh@um.ac.ir (J. Sadeh).}
}


%


\IEEEpubid{\makebox[\columnwidth]{978-1-7281-8192-9/21/\$31.00~\copyright2021 IEEE \hfill} \hspace{\columnsep}\makebox[\columnwidth]{ }}

\maketitle

\begin{abstract}
This paper is an effort to evaluate the numerical indices for the assessment of the frequency response analysis (FRA) to provide a deeper understanding of their characteristics. The study introduces the indices in the literature and categorizes them into two groups. The results of an actual FRA experimental setup on an electrical machine are employed to examine the indices from the various aspects. The main features of the indices are extracted, and the better ones in various aspects are marked. Moreover, based on the observations from the experimental studies, a new numerical index for comparing FRA results is proposed in this paper to enhance FRA results assessment and interpretation in both electrical machines and transformers. The advantages of the proposed index are also shown by comparing it with the existing ones. 
\end{abstract}

%
\IEEEpeerreviewmaketitle

\section{Introduction}

Frequency Response Analysis (FRA) is a comparison-based method for electrical apparatus condition assessment, which was introduced by Dick and Erven for transformers’ fault detection in 1978 \cite{2}. Since then, many research works are done in the field of transformers’ FRA. As a result of these researches and the FRA method's remarkable performance, it is currently known as one of the most reliable fault detection techniques, and organizations such as IEC, IEEE, and CIGRE standardized it as a routine test for transformers \cite{3}.

Electrical machines have always been the bed stone of industrial developments even in the emerging ones like autonomous vehicles \cite{k1}. Also, they are the main energy generation source in power systems besides non-rotational sources like PV systems \cite{armin} and batteries \cite{reza}. Hence, electrical machines' flawless operation is as important as system security \cite{ramin} for the grid operation. As electrical machines are very close in structure to the transformers, FRA recently is used as a fault detection method for electrical machines, and it is in its first steps of development. 

Due to the novelty of the FRA method for electrical machines, a few numbers of researches are done in this field. Most of the existing literature try to verify the application of this method in electrical machines by performing various FRA experiments. It has been shown that FRA can be used as a quality control method in the manufacturing of electrical machines \cite{20}. It can also detect stator problems of squirrel cage asynchronous machines \cite{20,21} or synchronous ones \cite{23}. Additionally, Stator's insulation evaluation is possible with FRA \cite{24,26}, and this test can also detect the synchronous machine's rotor faults \cite{28}. 

Further investigations are still required to analyze and verify the applicability of the FRA method in electrical machines like the evaluation of FRA application for a three-phase rotor fault diagnosis. Furthermore, assessment and interpretation of electrical machines FRA results are another important research area that needs investigation to expand the use of FRA in the electrical machines. Improvement in FRA results assessment also paves the way for the development of fault detection mechanisms based on the FRA with minimum human interface. In practice, it is possible to use numerical indices besides visual comparison of FRA results to enhance results assessments. Although several works evaluated the performance of numerical indices by applying them on actual or simulated transformers' FRA results \cite{8,9,E2}, few works dealt with this topic in the electrical machines field. In line with the works in stochastic modeling area \cite{zahra}, reference \cite{29} introduces a trend-based comparison method for enhancing assessments of electrical machines' FRA results.

This paper evaluates the performance of a large number of numerical indices for assessing FRA results of electrical machines. The evaluations are done by applying twenty-two numerical indices on actual FRA results measured from the stator of a three-phase electrical machine while various inter-turn short-circuit faults are implemented on it. To the best of our knowledge, this number of indices are not compared with each other using a single set of FRA results in any other FRA studies in either transformers or electrical machines. The most accurate and sensitive indices are marked, and the pros and cons of each index are mentioned by doing various analyses. More importantly, a new numerical index is proposed in this paper to assess FRA results, which has a remarkable performance compared to the existing ones and can enhance FRA diagnosis accuracy. 

The rest of the paper is arranged as follows. Section II provides an overview of the FRA test concept, introduces the test setup, and presents the FRA benchmark results for evaluation. Section III evaluates the performance of various indices, while Section IV proposes the new index for assessing FRA results. Finally, Section V concludes the paper. 

\section{Experimental FRA Test Setup}

\subsection{FRA Principle}

FRA is an offline non-invasive and non-destructive test based on measuring windings' response in a wide frequency range. The responses in the form of transfer functions are directly related to the device's RLC network. Geometry and construction of the windings determine the device’s RLC network, so any changes in the windings can affect the RLC network and also the frequency responses. Consequently, comparing the measured FRA test result with the reference response (signature or fingerprint), which is measured when the device is in sound condition, can reveal changes in the device’s condition and results in fault diagnosis \cite{3}.

For measuring the frequency response, a sweep frequency or impulse signal is injected into one of the winding’s terminals called the input signal. The output signal is then measured in any other terminals of the device according to the measurement type. Subsequently, the frequency response magnitude and phase angles are calculated using input and output signals concerning the intended transfer function. Due to the FRA method's comparative nature, the type of the used transfer function does not affect the test principle, while it may cause changes in the quality of the FRA results. For example, different transfer functions may have different accuracy in fault detection. Generally, the frequency response can be measured as a voltage ratio of two terminals or as an input impedance or admittance of a winding \cite{3}.

\subsection{Experimental Test Setup}

\begin{figure}[!h]
\centering
\includegraphics[width=3.4in]{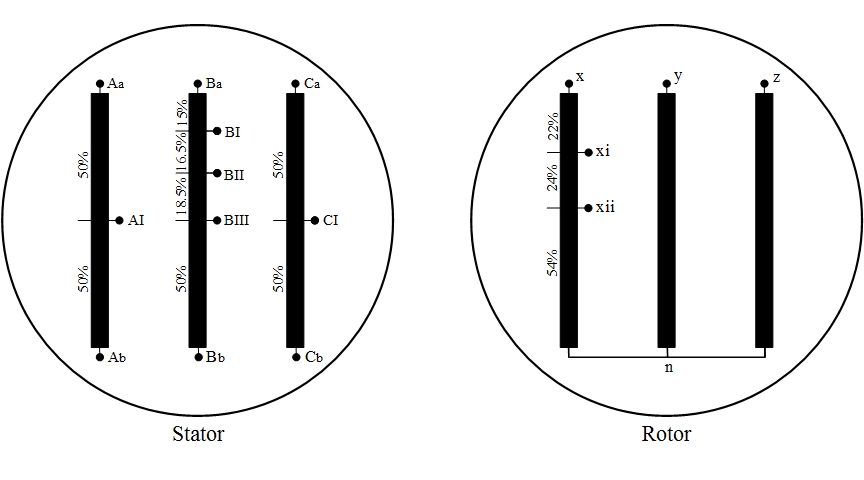}
\caption{Winding map of the machine used for doing experimental FRA tests}
\label{winding_map}
\end{figure}

Analyses of this paper are done based on actual FRA results measured from a three-phase electrical machine. A 4.5kW, 220/380V three-phase wound-rotor asynchronous machine is used for implementing faults and measuring the FRA results. The specific structure of this machine provides five terminals on different locations of the stator windings (terminals AI, BI, BII, BIII, and CI in Fig. \ref{winding_map}) and three terminals on the rotor windings (terminals xi, xii, and n in Fig. \ref{winding_map}). These terminals are separate from the standard terminals of this kind of machine (terminals Aa, Ab, Ba, Bb, Ca, Cb, x, y, and z in Fig. \ref{winding_map}) and are used for implementing faults. Fig. \ref{winding_map} demonstrates the machine’s windings' map. 

\begin{figure}[!h]
\centering
\includegraphics[width=3.2in]{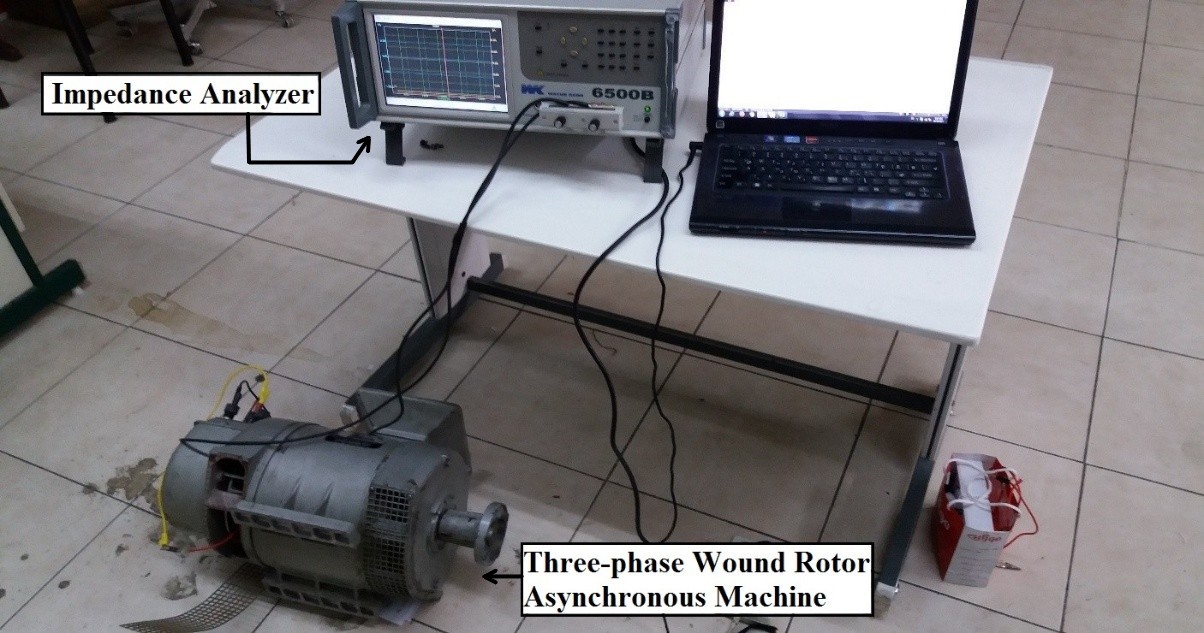}
\caption{Experimental test setup}
\label{test_setup}
\end{figure}

Input impedance is chosen as the transfer function for the FRA tests of this paper. Winding’s input impedance is measured over a frequency range by Wayne Kerr Electronics 6530B impedance analyzer to perform the FRA test. This equipment measures impedance ($Z$) and phase angle ($\theta$) in the 20 Hz to 30 MHz frequency range with a $\pm$0.05\% error. Fig. \ref{test_setup} shows the whole test setup of this work.

\subsection{Benchmark FRA Test}

\begin{figure}[!h]
\centering\includegraphics[width=3.4in]{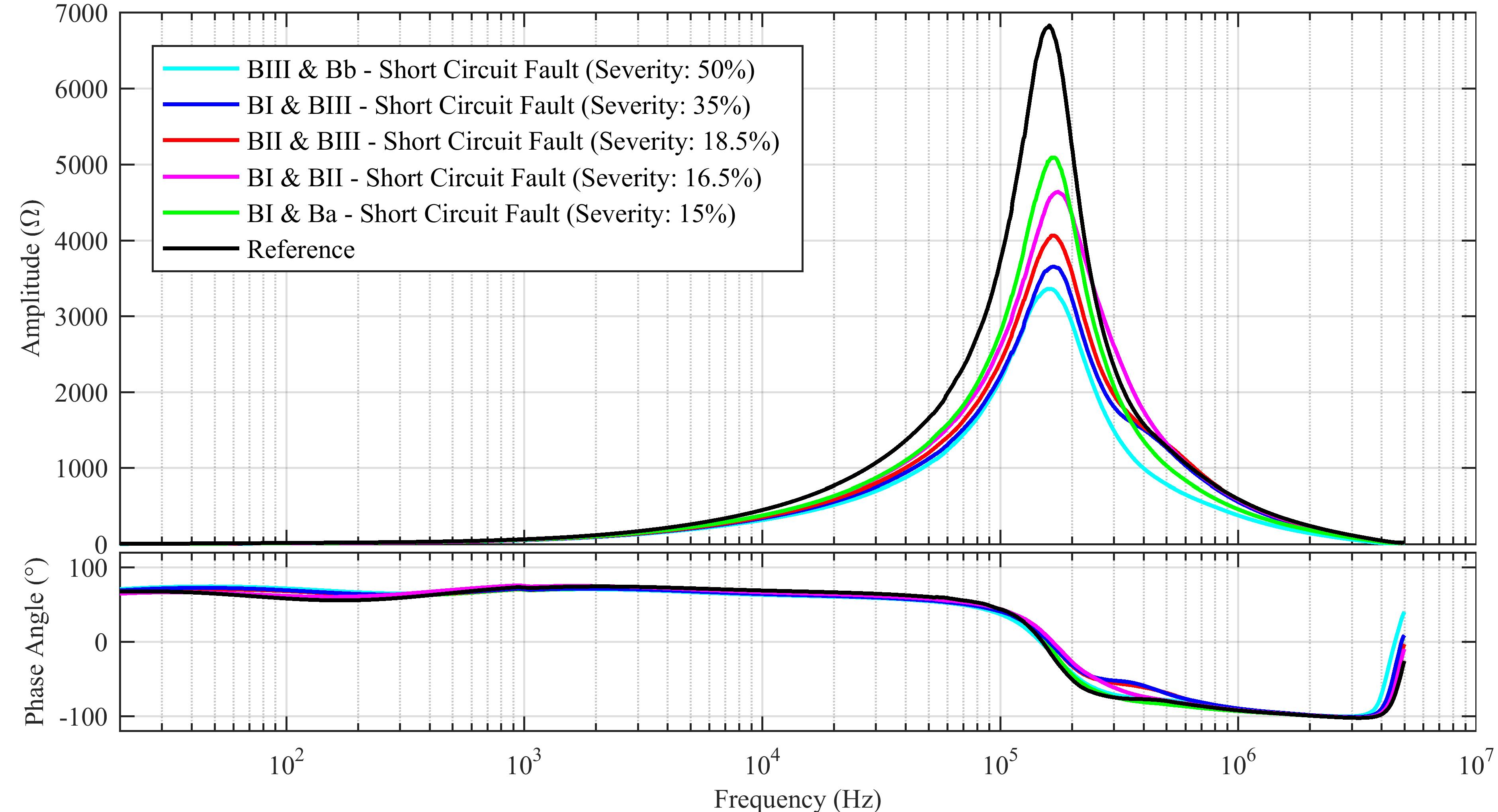}
\caption{Stator's phase B frequency response when inter-turn short-circuit fault is implemented on it in various locations}
\label{interfault}
\end{figure}

Several FRA tests are done on the above-mentioned electrical machine in faulty conditions to provide a benchmark for evaluating the numerical indices on it. The stator’s phase B impedance is acquired when inter-turn faults are implemented on different stator taps. The implemented faults are examples of real-life faults that may happen on electrical machines and show how FRA results change due to these faults. Hence, the measured FRA results are a benchmark for analyzing the performance of numerical indices in FRA results assessment. The amplitude and angle of these FRA results are shown in Fig. \ref{interfault} along with the FRA result of the stator in sound condition (black curve). The evaluation of numerical indices is done in the following sections by applying them to the amplitude diagram of this figure. Note that the amplitude diagram is the main resource for fault detection in FRA results, and phase angle may be used in some specific cases.
\section{Evaluation of Numerical Indices for FRA Assessment}

In this section, twenty-two numerical indices are applied to the experimental results to find the most accurate and sensitive ones. Also, each index's pros and cons are mentioned for better application of them in the FRA assessment. Understudy numerical indices can be categorized into two groups. The first group is one-array indices, which are calculated with one set of data, either reference curve or faulty curve. The average index is a well-known example of one-array indices. For comparing the faulty curve with the reference one with this type of indices, the index must be calculated for each curve separately, then, the difference between the calculated values for indices shows the degree of variation between compared curves. The second group of indices is two-array indices, which use both data sets simultaneously for calculation of the index. For these indices, the value of the index represents the difference between the curves. The correlation coefficient index is a good example of this group of indices. 

\subsection{Evaluation of One-array Indices }

The names, abbreviations, and equations of one-array indices are presented in Table \ref{one}. To use these indices, first, the value of each index for each faulty curve of Fig. \ref{interfault} is calculated. Second, it is calculated how much the index value for the faulty curve is changed with respect to those of the reference curve (in percentages). The change percentage is the indicator of variation between the faulty curve and reference one and can be used for assessing FRA results.

\begin{table}[!h]
\renewcommand{\arraystretch}{0.9}
\caption{One-array indices}
\label{one}
\centering
\begin{tabular}{|l|l|}
\hline
Index & Description \\
\hline
\multirow{2}{*}{Average ($Av$)} & \multirow{2}{*}{$Av=\Bar{x}=\frac{1}{N}\sum_{i=1}^{N}x_i$}\\ 
& \\
\multirow{2}{*}{Median ($M$)} & \multirow{2}{*}{$M=L+(\frac{(N/2)-cf}{f})h$} \\
 & \\
\multirow{2}{*}{Harmonic Mean ($Har$)} & \multirow{2}{*}{$\frac{1}{Har}=\frac{1}{N}\sum_{i=1}^{N}\frac{1}{x_i}$} \\
& \\
\multirow{2}{*}{Range ($R$)} & \multirow{2}{*}{$R=max(x)-min(x)$} \\
& \\
\multirow{2}{*}{Average Deviation ($AD$)} & \multirow{2}{*}{$AD=\frac{1}{N}\sum_{i=1}^{N}|x_i-\Bar{x}|$}\\
& \\
\multirow{2}{*}{Variance ($V$)} & \multirow{2}{*}{$V=\frac{1}{N}\sum_{i=1}^{N}(x_i-\Bar{x})^2$}\\
& \\
\multirow{2}{*}{Standard Deviation ($SD$)} & \multirow{2}{*}{$SD=\sqrt{\frac{1}{N}\sum_{i=1}^{N}(x_i-\Bar{x})^2}$}\\
& \\
\multirow{2}{*}{Standard Error of Mean ($SEM$)} & \multirow{2}{*}{$SEM=\frac{SD}{\sqrt{N}}$}\\
& \\
\multirow{2}{*}{Relation Dispersion ($RD$)} & \multirow{2}{*}{$RD=100(\frac{SD}{\Bar{x}})$}\\
& \\
\hline
\multicolumn{2}{|l|}{$N$: Size of the data set} \\
\multicolumn{2}{|l|}{$L$: Lower limit of the median class} \\
\multicolumn{2}{|l|}{$cf$: Cumulative frequency of classes prior to the median class} \\
\multicolumn{2}{|l|}{$f$: Frequency of median class} \\
\multicolumn{2}{|l|}{$h$: Median class size} \\
\multicolumn{2}{|l|}{$\Bar{x}$: Average value of $X$ data set} \\
\multicolumn{2}{|l|}{$x_i$: Data points of $X$ data set} \\
\hline
\end{tabular}
\end{table}

Instead of reporting the quantitative values of the indices' change percentages for each fault, calculated change percentages are normalized based on the index's greatest change percentage among the faulty curves of Fig. \ref{interfault}. Hence, for each index, each faulty curve index's difference relative to the reference curve index is divided over the highest difference for the calculated indices for the curves of Fig. \ref{interfault}. Equation \eqref{norm1} normalizes the average index as an example of one-array indices normalization process. In this equation, $x$ is the data array of the reference curve, $y$ is the data array of each faulty curves, and $Y$ is a set of all faulty curves of Fig. \ref{interfault}.
\begin{equation}
Av(y)_{Norm}=\frac{Av(x)-Av(y)}{\underset{\forall y \in Y}{Max}[Av(x)-Av(y)]}    
\label{norm1}
\end{equation}

\begin{figure}[!h]
\centering\includegraphics[width=3.4in]{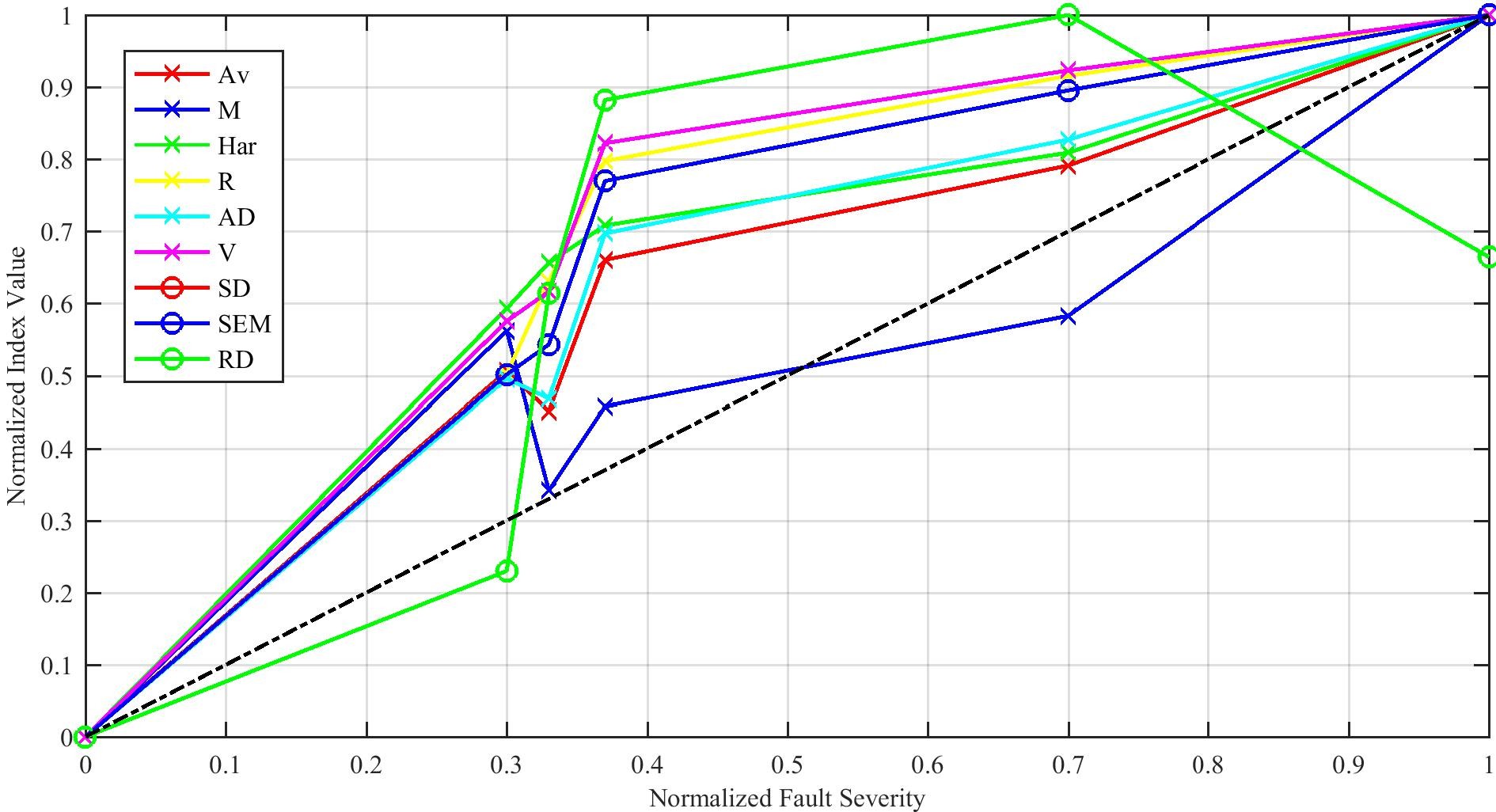}
\caption{Normalized change percentage of one-array indices in different short-circuit fault levels}
\label{one-arr}
\end{figure}

The calculated normalized values for the indices are plotted versus normalized faults' severity in Fig. \ref{one-arr} to evaluate indices and compare them. For normalizing faults' severity, the percentage of the short-circuited portion on winding in each fault is divided over 50, which is the most severe implemented fault (short-circuit between taps BIII and Ba in Fig. \ref{winding_map}).

In Fig. \ref{interfault}, as the fault becomes more severe, the difference between its curve and the reference curve increases. Although the difference between curves is not constant all over the frequency range and may overlap and cross each other in some parts, it is rational to expect that the visible increment of difference between faulty curves and reference be apparent in the calculated indices. Therefore, each index curve must have a monotone increment or decrement trend. According to Fig. \ref{one-arr}, normalized curves of \textit{Av}, \textit{M}, \textit{AD}, and \textit{RD} indices do not have a monotone trend, and in at least one stage, they decreases. This issue is a weak point for these indices as they are not able to follow the result changes accurately.

The index behavior as the fault severity change is another important issue. In other words, it is crucial that how proportionally index value changes relative to the changes in the fault severity. As mentioned, FRA results change relative to the fault severity, but these changes are not necessarily linear. Hence, completely proportional behavior is not expected in the percentage change of the indices. However, the index that changes more proportional to the fault severity change is better, especially for fault severity determination purposes. Among the indices that have a monotone trend in Fig. \ref{one-arr}, \textit{Har} index curve is the closest one to the dotted line (perfectly proportional behavior concerning fault severity changes), and it has the most proportional behavior to fault severity changes.

The sensitivity and accuracy of the indices are also evaluated over the faults with close severity levels. In Fig. \ref{interfault}, the least severe implemented fault is the green curve, which short-circuits 15\% of the winding (fault level one). The second severe fault is the purple curve, which short-circuits 16.5\% of the winding (fault level two). The fault level three (red curve) is two percent more severe than the second one (18.5\% short-circuited). The severity levels of these faults change 1.5\%, 2\%, and 3.5\% relative to each other, which are very close. 

For each of the indices, the relative two-by-two absolute change percentage of the index values over these three fault levels are calculated. It means that for each index, it is calculated that how much its value for level two fault changes relative to level one; and how much its value changes for level three fault relative to level two fault. Similarly, the change percentage of index value for level three fault relative to level one is measured. The average of these change percentages is presented in Table \ref{sen-one} as an indicator of the indices' accuracy. The index with a higher average is more sensitive to the small changes in faults' severity.
\begin{table}[!h]
\caption{Average of two-by-two change percentage of one-array index values for first three fault severity levels}
\label{sen-one}
\centering
\begin{tabular}{|c||c|c|c|c|c|}
\hline
 Index & \textit{M} & \textit{Av} & \textit{Har} & \textit{R} & \textit{AD} \\
\hline
Change Percentage & 30.5 & 29.2 & 12.6 & 36.9 & 31.6 \\ 
\hline
\hline
Index & \textit{V} & \textit{SD} & \textit{SEM} & \multicolumn{2}{c|}{\textit{RD}} \\
\hline
Change Percentage & 27.7 & 34.5 & 34.5 & \multicolumn{2}{c|}{164.7} \\ 
\hline
\end{tabular}
\end{table}

According to Table \ref{sen-one}, the \textit{RD} index has the most significant value, but it does not have a monotone trend curve and decreases in the last interval. Hence, this index is sensitive and accurate for detecting small faults, but it is not reliable in more severe faults. Among other indices that have approximately the same sensitivity, the \textit{V} index is more accurate for fault detection as its normalized curve is above all curves in Fig. \ref{one-arr}.

\subsection{Evaluation of Two-array Indices}

Table \ref{two} shows names, abbreviations, and equations of two-array indices. For evaluating two-array indices, their values are calculated over the curves of Fig. \ref{interfault}, and similar to the previous section, they are normalized and depicted versus normalized values of faults' severity. Note that for two-array indices, the value of the index itself is a measure for the difference between compared curves. Also, the normalization procedure for these indices is not always the same as one-array ones.

\begin{table*}[t]
\renewcommand{\arraystretch}{0.9}
\caption{Two-array indices}
\label{two}
\centering
\begin{tabular}{|l|l|}
\hline
Index & Description \\
\hline
\multirow{2}{*}{Correlation Coefficient ($CC$)} & \multirow{2}{*}{$CC=\frac{\sum_{i=1}^N(x_i-\Bar{x})(y_i-\Bar{y})}{\sqrt{\sum_{i=1}^N(x_i-\Bar{x})^2(y_i-\Bar{y})^2}}$}\\ 
& \\
\multirow{2}{*}{Covariance ($COV$)} & \multirow{2}{*}{$COV=\frac{1}{N}\sum_{i=1}^N(x_i-\Bar{x})(y_i-\Bar{y})$} \\
& \\
\multirow{2}{*}{Absolute Sum of Logarithmic Error ($ASLE$)} & \multirow{2}{*}{$ASLE=\frac{\sum_{i=1}^N|20\log|x_i|-20\log|y_i||}{N}$} \\
& \\
\multirow{2}{*}{Absolute Average Difference ($DABS$)} & \multirow{2}{*}{$DABS=\frac{1}{N}\sum_{i=1}^{N}|x_i-y_i|$} \\
& \\
\multirow{2}{*}{Min-Max Index ($MM$)} & \multirow{2}{*}{$MM=\frac{\sum_{i=1}^{N}min(x_i,y_i)}{\sum_{i=1}^{N}max(x_i,y_i)}$} \\
& \\
\multirow{2}{*}{Sum of Squared Error ($SSE$)} & \multirow{2}{*}{$SSE=\frac{1}{N}\sum_{i=1}^{N}(x_i-y_i)^2$}\\
& \\
\multirow{2}{*}{Root of Sum of Squared Error ($RSSE$)} & \multirow{2}{*}{$RSSE=\sqrt{\frac{1}{N}\sum_{i=1}^{N}(x_i-y_i)^2}$}\\
& \\
\multirow{2}{*}{Sum of Squared Ratio Error ($SSRE$)} & \multirow{2}{*}{$SSRE=\frac{1}{N}\sum_{i=1}^N(\frac{y_i}{x_i}-1)^2$}\\
& \\
\multirow{2}{*}{Sum of Squared Max-Min Ratio Error ($SSMMRE$) \cite{30}} & \multirow{2}{*}{$SSMMRE=\frac{1}{N}\sum_{i=1}^N(\frac{max(x_i,y_i)}{min(x_i,y_i)}-1)$}\\
& \\
\multirow{2}{*}{Comparative Standard Deviation ($CSD$)} & \multirow{2}{*}{$CSD=\sqrt{\frac{1}{N}\sum_{i=1}^N[(x_i-\Bar{x})-(y_i-\Bar{y})]^2}$}\\
& \\
\multirow{2}{*}{Spectrum Deviation ($SpD$)} & \multirow{2}{*}{$SpD=\frac{1}{N}\sum_{i=1}^N\sqrt{(\frac{x_i}{(x_i+y_i)/2}-1)^2+(\frac{y_i}{(x_i+y_i)/2}-1)^2}$} \\
& \\
\multirow{3}{*}{Standardized Difference Area ($SDA$)} & \multirow{3}{*}{$SDA=\frac{\int_{f_{min}}^{f_{max}}|x(f)-y(f)|df}{\int_{f_{min}}^{f_{max}}x(f)df}$} \\
& \\
& \\
\multirow{2}{*}{Euclidean Distance ($ED$)} & \multirow{2}{*}{$ED=\sqrt{\sum_{i=1}^{N}(x_i-y_i)^2}$}\\
& \\
\hline
\multicolumn{2}{|l|}{$N$: Size of the data set} \\
\multicolumn{2}{|l|}{$f$: Frequency} \\
\multicolumn{2}{|l|}{$\Bar{x}, \Bar{y}$: Average value of $X$ and $Y$ data sets} \\
\multicolumn{2}{|l|}{$x_i$, $y_i$: Data points of $X$ and $Y$ data sets} \\
\hline
\end{tabular}
\end{table*}

The values of the two-array indices except for \textit{CC}, \textit{COV}, \textit{MM} grow as dissimilarity of compared curves increases. Thus, for normalization, each curve's index value is divided over the highest index value for curves of Fig. \ref{interfault}. For instance, equation \eqref{norm2-1} normalize the \textit{DABS} index. In this equation, $x$ is the data array of the reference curve, $y$ is the data array of each faulty curves, and $Y$ is a set of all faulty curves of Fig. \ref{interfault}.
\begin{equation}
    DABS(x,y)_{norm}=\frac{DABS(x,y)}{\underset{\forall y \in Y}{Max}[DABS(x,y)]}
    \label{norm2-1}
\end{equation}

The \textit{CC}, \textit{COV}, and \textit{MM} indices do not behave like other ones, e.g., \textit{CC} and \textit{MM} indices' values are one when two curves are identical and decrease as the curves become more different from each other. On the other hand, \textit{COV} values increase as the curves get more similar to each other. Hence, for normalizing these indices, first, the index is calculated for the sound condition (reference curve is compared with itself, which results in a large number for \textit{COV} and one for \textit{CC} and \textit{MM}). Then, the difference of index values of each faulty curves relative to the index value of sound condition is divided over the highest difference. Equation \eqref{norm2-2} normalizes the MM index as an example. In this equation, $x$ is the data array of the reference curve, $y$ is the data array of each faulty curves, and $Y$ is a set of all faulty curves of Fig. \ref{interfault}.
\begin{equation}
    MM(x,y)_{norm}=\frac{MM(x,x)-MM(x,y)}{\underset{\forall y \in Y}{Max}[MM(x,x)-MM(x,y)]}
    \label{norm2-2}
\end{equation}

\begin{figure}[!h]
\centering
\includegraphics[width=3.4in]{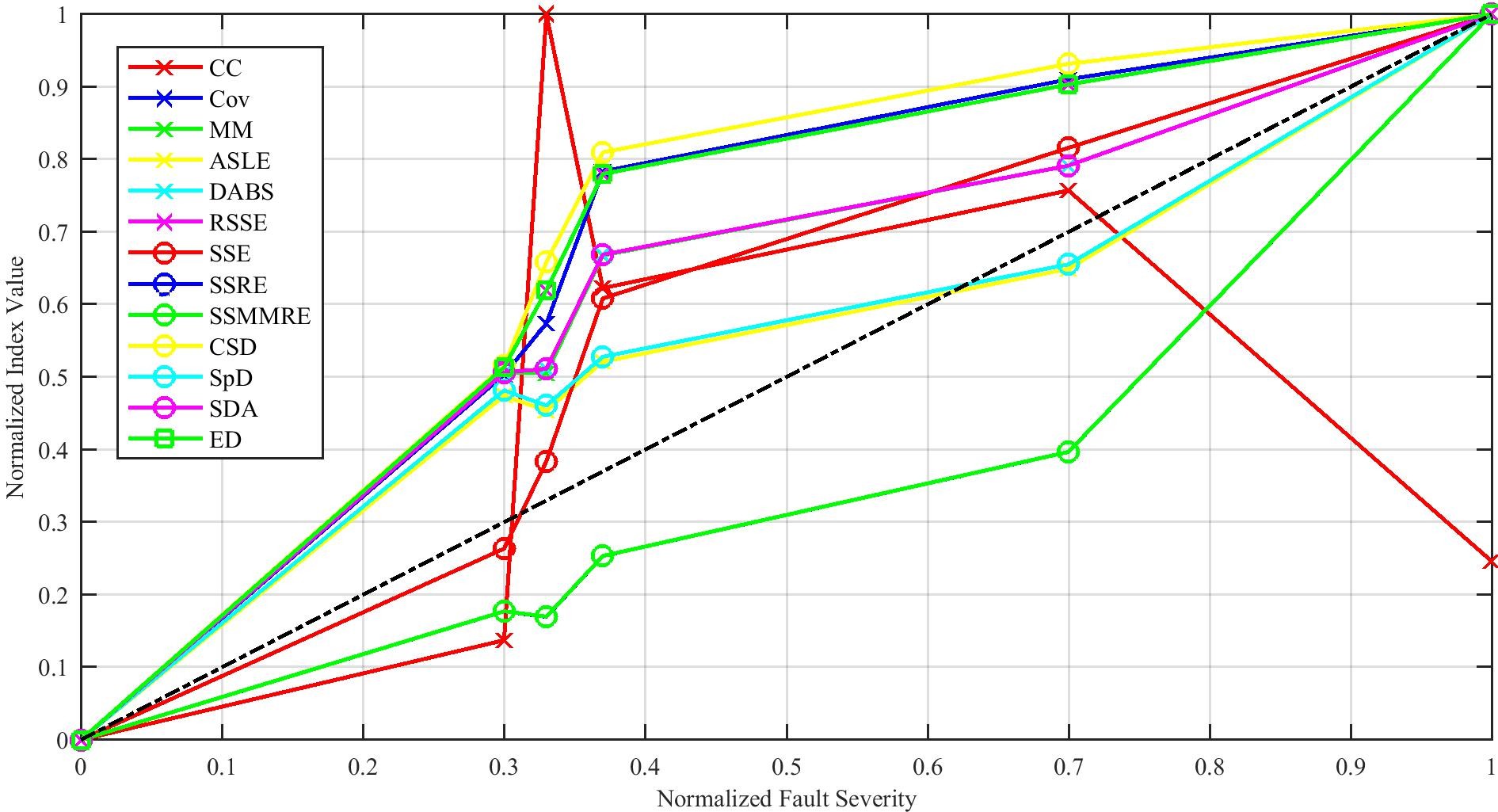}
\caption{Normalized change percentage of two-array indices in different short-circuit fault levels}
\label{two-arr}
\end{figure}

Fig. \ref{two-arr} shows the normalized values of two-array indices versus normalized faults' severity. The index value of all two-array indices except \textit{CC}, \textit{COV}, and \textit{MM} grows as the compared curves become more apart from each other, and they must have monotone increasing curves. Additionally, although the index value of \textit{CC}, \textit{COV}, and \textit{MM} decreases by fault expansion; due to the described normalization procedure for these three indices, the normalized curves of them must also have an increasing trend. Fig. \ref{two-arr} shows that \textit{ASLE}, \textit{SSRE}, \textit{SSMMRE}, and \textit{SpD} indices do not have a monotone trend, and in some fault levels, curves' trend changes. 

Two-array indices change more proportionally to the severity changes than one-array indices, and their curves are closer to the dotted reference line of perfectly proportional behavior. It is apparent in Fig. \ref{two-arr} that \textit{ASLE} and \textit{SpD} have the most proportional behavior to the fault severity changes than all of the one- and two-array indices. It must be noted that these two indices do not have a monotone trend, which must be considered in their application. Among the indices that have monotone trend, \textit{DABS}, \textit{SDA}, and \textit{SSE} are closer to the dotted line and are more linear or behave more proportionally to fault severity changes. Note that the normalized curve of \textit{DABS} and \textit{SDA} are similar to each other and overlap. 

The sensitivity of two-array indices is also analyzed with the same approach as the previous section. Average of two-by-two change percentage of index values are calculated for the first three fault levels and reported in Table \ref{sen-two}. Based on this table, \textit{SSE} is the most sensitive index that, on average, changes about 78\% for, on average, two percent of fault severity change. Fig. \ref{two-arr} also shows that \textit{SSE} sharply increases in the first three fault levels, but it does not change like that for more severe faults. Hence, The \textit{SSE} index can be more useful for small difference detection in the FRA results.

\begin{table}[!h]
\caption{Average of two-by-two change percentage of two-array index values for first three fault severity levels}
\label{sen-two}
\centering
\begin{tabular}{|c||c|c|c|c|c|}
\hline
 Index & \textit{CC} & \textit{COV} & \textit{MM} & \textit{ASLE} & \textit{DABS} \\
\hline
Change Percentage & 1.2 & 11.3 & 5.5 & 9.7 & 21.2  \\ 
\hline
\hline
Index & \textit{SSE}& \textit{RSSE} & \textit{CSD} & \textit{SpD} & \textit{SDA}  \\
\hline
Change Percentage & 78.3 & 32.8 & 35.9 & 9.5 & 21.2 \\ 
\hline
\hline
Index & \textit{ED} & \multicolumn{2}{c|}{\textit{SSRE}} & \multicolumn{2}{c|}{\textit{SSMMRE}} \\
\hline
Change Percentage &  32.8 & \multicolumn{2}{c|}{32.2} & \multicolumn{2}{c|}{32.1} \\ 
\hline
\end{tabular}
\end{table}

\section{A New Index for FRA Results Assessment}

The evaluations of the previous section showed that the \textit{DABS} index has a good performance in FRA results assessments. In the preformed case studies, first, the \textit{DABS} index curve in Fig. \ref{two-arr} has a monotone increasing trend as the fault severity increases. Second, this index is among the indices whose value change more proportional to the fault severity changes. Third, it has an acceptable sensitivity in comparison to other indices (Table \ref{sen-two}). The good performance of the \textit{DABS} index for the FRA result assessment is endorsed in other researches in this field \cite{8}. According to Table \ref{two}, the \textit{DABS} index calculates the average of the absolute difference between the compared curves or data sets. This calculated value is more meaningful if someone considers the range of curve's magnitude change. For example, if an FRA curve’s magnitude range is 20 and has \textit{DABS} index value of five, and another FRA curve’s range is 50 but has the same \textit{DABS} value; the former FRA result belongs to a more severe fault as its average absolute difference with the reference curve is a big number in comparison to its range. 

As a result, in this paper, the ratio of \textit{DABS} index to the $R$ (Range) of the faulty curve (or any other curve that is compared with reference) is proposed as a new index for assessing FRA results. This index, which can be called the Absolute Average Difference to Range Ratio (\textit{AADRR}), is calculated by \eqref{AADRR}.
\begin{equation}
    AADRR=\frac{DABS(x,y)}{R(y)}=\frac{\frac{1}{N}\sum_{i=1}^{N}|x_i-y_i|}{max(y)-min(y)}
    \label{AADRR}
\end{equation}
where $x$ is the array of reference curve data points, $y$ is the faulty curve data set, and $N$ is the size of these sets.

\begin{figure}[t]
\centering\includegraphics[width=3.35in]{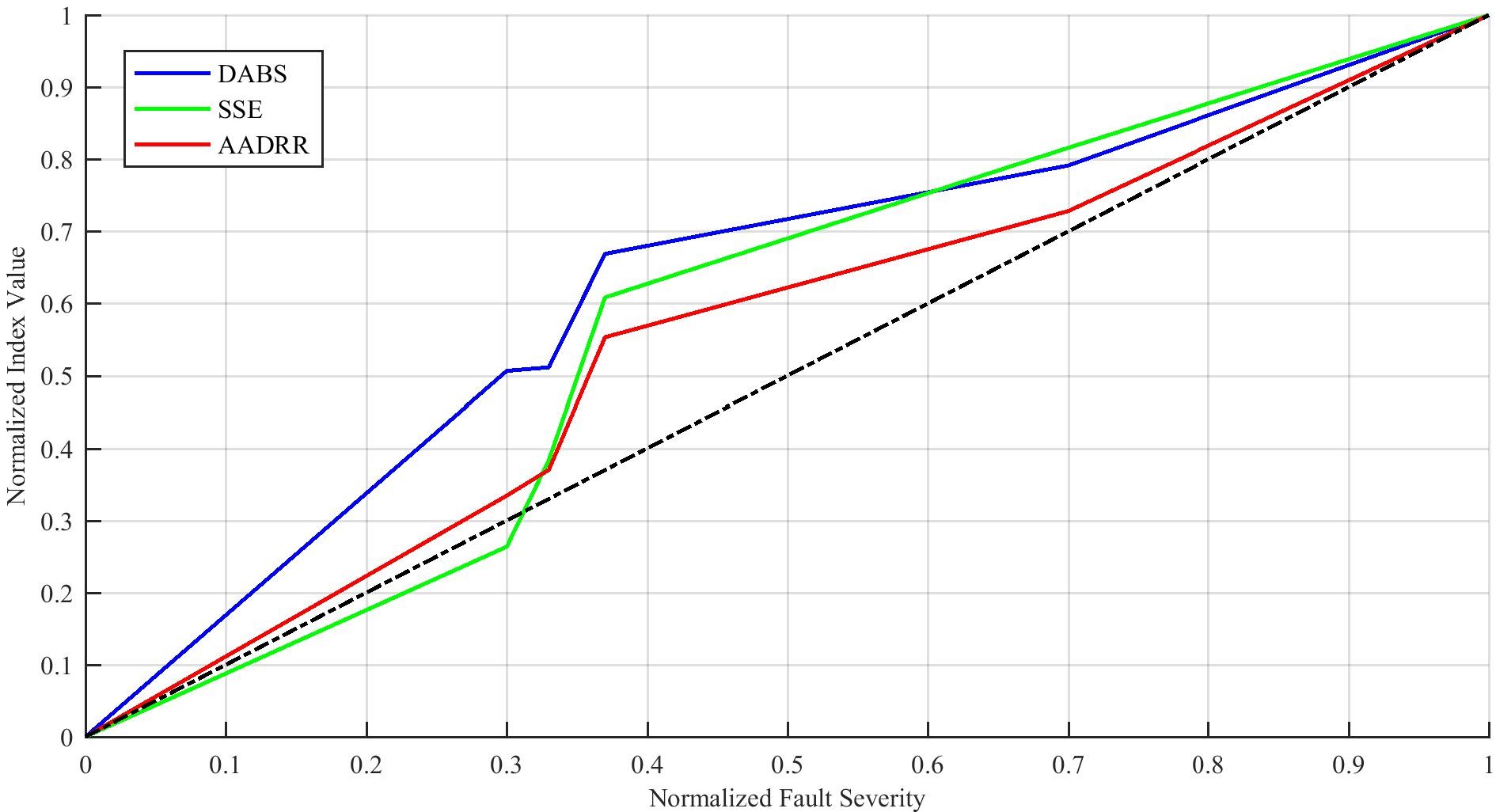}
\caption{Normalized curves of AADRR, DABS and SSE indices}
\label{new-index}
\end{figure}

The \textit{AADRR} index is applied to the curves of Fig. \ref{interfault} to evaluate its performance. Its normalized curve is derived with the same procedure described before (equation \eqref{norm2-1}), and plotted in Fig. \ref{new-index}. In this figure, the normalized curve of \textit{AADRR} is compared with \textit{DABS} and \textit{SSE}, which were among the best in the previous section evaluation. According to Fig. \ref{new-index}, first, the \textit{AADRR} index has a monotone increasing trend and entirely follows the changes of faults' severity. Second, this index changes more proportional to fault severity than all other numerical indices that are studied in this work.

In Fig. \ref{interfault} and most of the FRA results \cite{3}, the range of curves ($R$) decreases as the fault severity increases. The \textit{AADRR} index must be more sensitive than the \textit{DABS} index as its denominator decreases when the fault severity grows. To size up \textit{AADRR} sensitivity, the average of two-by-two change percentage of this index for the first three fault levels is calculated, which is equal to 41.8 \%. It is a considerable number and is just less than the \textit{SSE} and \textit{RD} indices. Based on the performed analysis on real experimental FRA results, the \textit{AADRR} index has an excellent performance.

\section{Conclusion}

In line with the researches focused on the FRA results' assessment, and in order to expand the FRA application in electrical machine, this paper provides a comprehensive evaluation of a large number of numerical indices, which are used for the FRA results assessment. These evaluations are done on actual FRA tests measured from a three-phase wound-rotor asynchronous machine while different inter-turn short-circuit faults are implemented on its stator. 

In the performed case studies, $V$ (variance) and \textit{RD} (relation dispersion) are the more sensitive one-array indices and \textit{Har} (Harmonic Mean) is the one that changes more proportional to the fault severity changes. Two-array indices in overall have better performance than one-array ones. Among them, \textit{SSE} (sum of squared error), \textit{SDA} (standardized difference area), and \textit{DABS} (absolute average difference) are more accurate and behave more proportionally to the fault severity changes. We also proposed a new index for FRA results assessment, which is called the absolute average difference to range ratio (\textit{AADRR}). Initial analysis showed that this index has a remarkably good performance. Hence, the \textit{AADRR} index has the potential to significantly enhance FRA results assessment in electrical machines or even transformers due to the similarities between these two devices' FRA results. 

Future works can be focused on using the new index and those with better performance for developing automatic detection procedures. Also, more practical and mathematical analyses must be done on the \textit{AADRR} index.

\bibliographystyle{IEEEtran}
\bibliography{IEEEabrv,ref}




\end{document}